\documentclass[reprint,amsmath,amssymb,aps,prb,
floatfix]{revtex4-2}

\usepackage[utf8]{inputenc}
\usepackage{xcite}
\usepackage{xr}
\usepackage{amsmath}
\usepackage{graphicx}
\usepackage[hidelinks]{hyperref}
\usepackage[margin=1in]{geometry}
\usepackage{lineno}
\usepackage{xcolor}

\begin{document}

\title{Role of substrate clamping on anisotropy and domain structure in the canted antiferromagnet $\alpha$-Fe$_2$O$_3$}

\author{Angela Wittmann$^{1,2}$}
\author{Olena Gomonay$^2$}
\author{Kai Litzius$^{1,3}$}
\author{Allison Kaczmarek$^{1}$}
\author{Alexander E. Kossak$^{1}$}
\author{Daniel Wolf$^{4}$}
\author{Axel Lubk$^{4}$}
\author{Tyler N. Johnson$^{5}$}
\author{Elizaveta A. Tremsina$^{1,6}$}
\author{Alexandra Churikova$^1$}
\author{Felix B\"uttner$^7$}
\author{Sebastian Wintz$^{3,7}$}
\author{Mohamad-Assaad Mawass$^{7}$}
\author{Markus Weigand$^{7}$}
\author{Florian Kronast$^{7}$}
\author{Larry Scipioni$^8$}
\author{Adam Shepard$^8$}
\author{Ty Newhouse-Illige$^8$}
\author{James A Greer$^8$}
\author{Gisela Sch\"utz$^3$}
\author{Norman O. Birge$^{1,9}$}
\author{Geoffrey S. D. Beach$^1$}

\affiliation{$^1$Department of Materials Science and Engineering, Massachusetts Institute of Technology, Cambridge, Massachusetts 02139, USA}
\affiliation{$^2$Institute of Physics, Johannes Gutenberg-University Mainz, 55128 Mainz, Germany}
\affiliation{$^3$Max Planck Institute for Intelligent Systems, 70569 Stuttgart, Germany}
\affiliation{$^4$Leibniz IFW Dresden, 01069 Dresden, Germany}
\affiliation{$^5$Department of Chemical Engineering and Materials Science, Michigan State University, East Lansing, Michigan 48824, USA}
\affiliation{$^6$Department of Electrical Engineering and Computer Science, Massachusetts Institute of Technology, Cambridge, Massachusetts 02139, USA}
\affiliation{$^7$Helmholtz-Zentrum Berlin f\"ur Materialien und Energie GmbH, 14109 Berlin, Germany}
\affiliation{$^8$PVD Products, Wilmington, Massachusetts 01887, USA}
\affiliation{$^9$Department of Physics and Astronomy, Michigan State University, East Lansing, Michigan 48824, USA}

\date{\today}

\begin{abstract}
Antiferromagnets have recently been propelled to the forefront of spintronics by their high potential for revolutionizing memory technologies. For this, understanding the formation and driving mechanisms of the domain structure is paramount.
In this work, we investigate the domain structure in a thin-film canted antiferromagnet $\alpha$-Fe$_2$O$_3$.
We find that the internal destressing fields driving the formation of domains do not follow the crystal symmetry of $\alpha$-Fe$_2$O$_3$, but fluctuate due to substrate clamping. This leads to an overall isotropic distribution of the N\'eel order with locally varying effective anisotropy in antiferromagnetic thin films. 
Furthermore, we show that the weak ferromagnetic nature of $\alpha$-Fe$_2$O$_3$ leads to a qualitatively different dependence on magnetic field compared to collinear antiferromagnets such as NiO.
The insights gained from our work serve as a foundation for further studies of electrical and optical manipulation of the domain structure of antiferromagnetic thin films.

\end{abstract}

\maketitle
\section{Introduction}
Antiferromagnets (AFMs) exhibit highly favorable properties such as robustness to external magnetic field, no stray field, and ultra-fast switching in the terahertz regime making them promising candidates for novel memory technologies~\cite{Olejnik2018,Jungwirth2016,Baltz2018,Jungwirth2018,Zelezny2018a}. Recent interest in this field, and the canted antiferromagnet $\alpha$-Fe$_2$O$_3$ in particular, was sparked by the experimental demonstration of electrical control and readout of the N\'eel order in metallic AFMs~\cite{Wadley2016, Bodnar2018, Godinho2018}, insulating AFM/heavy-metal bilayers~\cite{Chen2018,Gray2019, Baldrati2019,Cheng2020,Zhang2019, Chiang2019, Churikova2020}, and the observation of topological spin textures~\cite{Chmiel2018,Jani2021a,Ross2020c} paving the path towards using AFMs as active materials for spintronic devices.

Analogously to demagnetizing fields in ferromagnets, destressing fields have been identified as the dominant driving force behind domain formation in AFMs~\cite{Gomonay2007,Gomonay2014}. So far, it was assumed that the domains are oriented along the magnetocrystalline easy-axes. However, the underlying model is strictly true only for free layers. Our experimental findings on $\alpha$-Fe$_2$O$_3$ thin films imply that substrate clamping results in a far more complex and disordered domain structure. We show that this leads to long-range interactions that impose a locally-varying effective anisotropy. The local fluctuations of the easy-axis orientations qualitatively impact the domain structure and magnetization process.
These previously unidentified effects should generally be present and in competition with the magnetocrystalline anisotropy in  antiferromagnetic oxide thin films based on their similarities in magnetic, chemical and structural properties~\cite{Rao1989,Duo2010}. Therefore, the insights and the refined model developed in this work will allow for a more informed understanding of magnetic switching and complex spin textures in easy-plane antiferromagnets.

\begin{figure*}[ht]
\centering\includegraphics[width=0.99\linewidth]{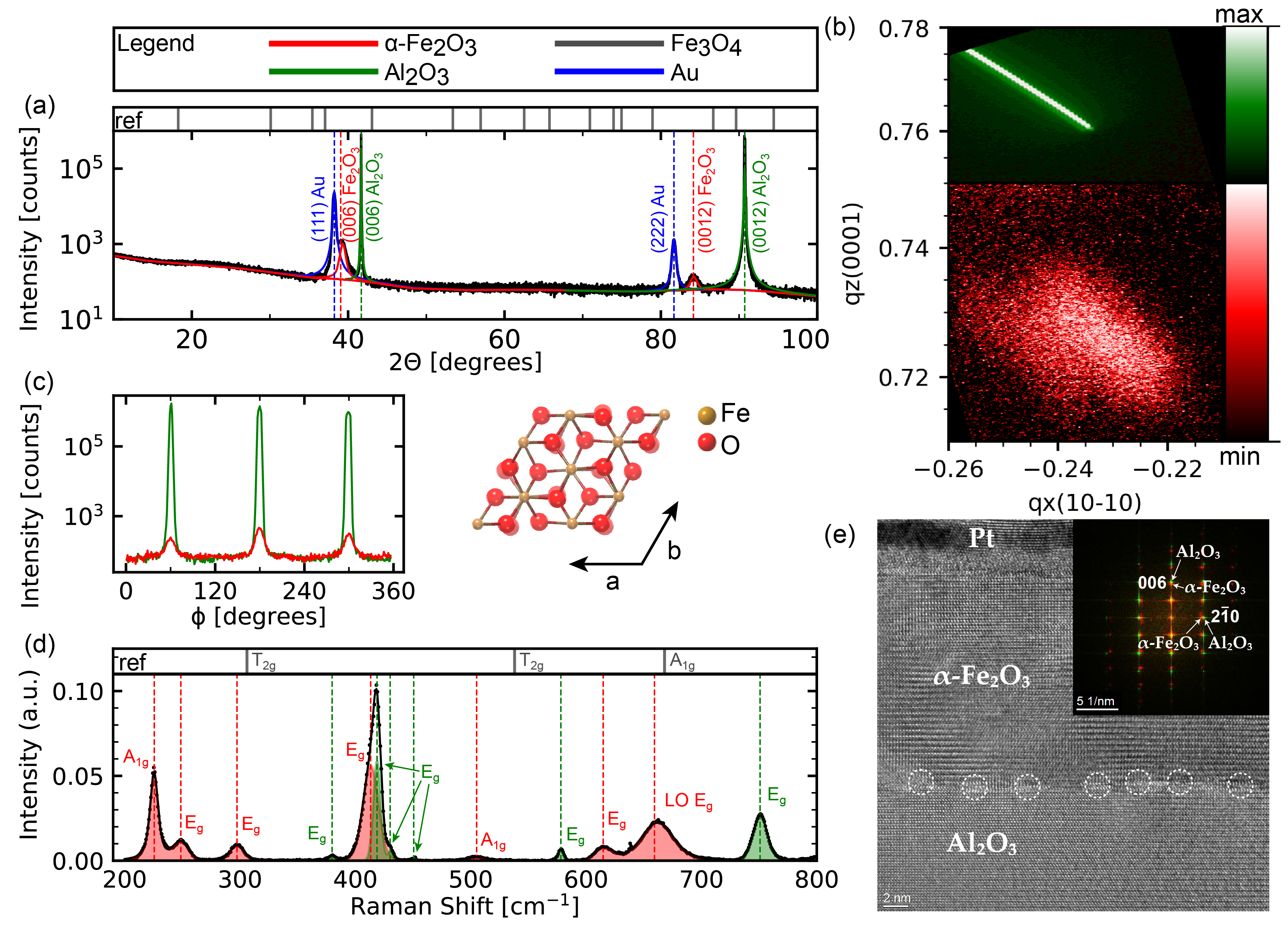}
\caption{(a) Long-range (006)/(0012) symmetric scan of $\alpha$-Fe$_2$O$_3$ (10~nm) on c-axis (001) oriented Al$_2$O$_3$. (b) (1 0 10)- reciprocal space map of $\alpha$-Fe$_2$O$_3$ (10~nm) on c-axis (001) oriented Al$_2$O$_3$. (c) Phi-scan of $\alpha$-Fe$_2$O$_3$ (20~nm) on c-axis (001) oriented Al$_2$O$_3$ with inset of hexagonal crystal structure. (d) Raman spectrum of $\alpha$-Fe$_2$O$_3$ (10~nm) thin film on a Al$_2$O$_3$ substrate. (e)~Cross-sectional HRTEM image of $\alpha$-Fe$_2$O$_3$ (20~nm) on c-axis (001) oriented Al$_2$O$_3$. The inset shows the Fourier transform of the image.}
\label{fig:1}
\end{figure*}

\section{Thin film characterization}
Our epitaxial $\alpha$-Fe$_2$O$_3$ films (10~nm thick unless mentioned otherwise) were grown at 640~$^\circ$C on $c$-axis (001) oriented Al$_2$O$_3$ with off-axis magnetron sputtering. In order to validate the quality of the thin films, we have performed structure and composition characterization.

Figure~\ref{fig:1}a-c show the structure characterization of the $\alpha$-Fe$_2$O$_3$ thin films using x-ray diffraction (XRD) confirming the high epitaxial quality of the thin films (see supplemental material for details~\cite{SI}).
The (006)/(0012) symmetric 2$\theta$ scan shown in Fig.~\ref{fig:1}a was used to determine the c lattice parameters of the magnetic thin film and the substrate (c$_{\alpha -Fe_2O_3}$ = 13.79~\AA\ and c$_{Al_2O_3}$ = 12.99~\AA). The film thickness was confirmed to be 11.88~nm (nominally 10~nm) using the Scherrer method which relates peak broadening to the crystallite size. The peaks from gold appear due to the gold contacts on the sample.
The strain state and degree of relaxation was determined from a reflection with in-plane and out-of-plane components. In the (1 0 10)- reciprocal space map (Fig.~\ref{fig:1}b),  shifting of the $\alpha$-Fe$_2$O$_3$ film peak around q$_x\approx -0.234$ (red) towards the origin is indicative of film relaxation with respect to the substrate at q$_x\approx -0.247$ (green). At the substrate-film interface, the film experiences strain due to clamping on the substrate, as shown by the shadow peak located along q$_x\approx -0.247$. Through the thickness of the $\alpha$-Fe$_2$O$_3$ film, the strain relaxes by 93.7~\% to a nearly fully relaxed film.
The phi-scan (Fig.~\ref{fig:1}c) confirms the six-fold symmetry in the $\alpha$-Fe$_2$O$_3$ and Al$_2$O$_3$ lattices.

To characterize the film-substrate interface in more detail, we acquired cross-sectional high-resolution transmission electron microscopy (HRTEM) images on a 20~nm $\alpha$-Fe$_2$O$_3$ film (see supplemental material for details~\cite{SI}). Fig.~\ref{fig:1}e shows an HRTEM image exhibiting clear atomic lattice plane contrast. The Fourier transform shown as inset in Fig.~\ref{fig:1}e confirms the high crystallinity of the $\alpha$-Fe$_2$O$_3$ film. To quantify the strain distribution along the thickness of the $\alpha$-Fe$_2$O$_3$ film, we conducted a geometric phase analysis~\cite{Hytch2011}. We find a sharp relaxation after 1~nm of the $\alpha$-Fe$_2$O$_3$ lattice clamped at the Al$_2$O$_3$ interface. This is facilitated by the presence of misfit dislocations and other crystal defects indicated with circles in the HRTEM image in Fig.~\ref{fig:1}e.
Moreover, we observe an almost constant in- and out-of-plane dilation in the $\alpha$-Fe$_2$O$_3$ lattice with respect to the smaller Al$_2$O$_3$ substrate lattice. The relative dilation by 6~\% agrees well with the ratio of the lattice plane distances for in-plane and out-of-plane directions. The sharp relaxation and dilation of the $\alpha$-Fe$_2$O$_3$ film support the findings from the XRD measurements described above.

Figure~\ref{fig:1}d shows the Raman measurement data (black dots) taken using an excitation laser wavelength of 532~nm with fits using Voigt functions (solid line). The center of the peaks are labelled by vertical dashed lines corresponding to the identified Raman modes based on previous reports in literature~\cite{Lubbe2010, Cvejic2009InfluenceMagnetite, Jubb2010VibrationalDeposition, Massey1990a, Porto1967RamanCorundum} (see supplemental material for detailed peak information~\cite{SI}).
In particular, we note the presence of a wide peak at 660~cm$^{-1}$ corresponding to the Raman-forbidden longitudinal optical (LO) mode in $\alpha$-Fe$_2$O$_3$ due to defect-induced scattering~\cite{Jubb2010VibrationalDeposition,Chernyshova2007Size-dependentTransition, Xu2009SynthesisNanoleaves}. 
We find that the experimental data is accurately fitted with Raman peaks corresponding to $\alpha$-Fe$_2$O$_3$ and Al$_2$O$_3$, with a small shift of up to 6~cm$^{-1}$ for $\alpha$-Fe$_2$O$_3$ compared to the bulk values. This shift agrees well with previous reports that compressive strain as observed in our in $\alpha$-Fe$_2$O$_3$ films in the XRD measurements leads to higher peak center values~\cite{Lubbe2010, Angel2019StressShifts}.\\

We note that the structure and composition characterization based on XRD, HRTEM, and Raman measurements presented here confirm the high epitaxial quality and phase-purity of the $\alpha$-Fe$_2$O$_3$ thin films grown on Al$_2$O$_3$ substrates investigated in this work. Importantly, we do not observe any evidence of impurity phases of Fe$_3$O$_4$ (see reference marks in Fig.~\ref{fig:1}a,d).

\section{Experiment}
\begin{figure}[hb]
\centering\includegraphics[width=0.99\linewidth]{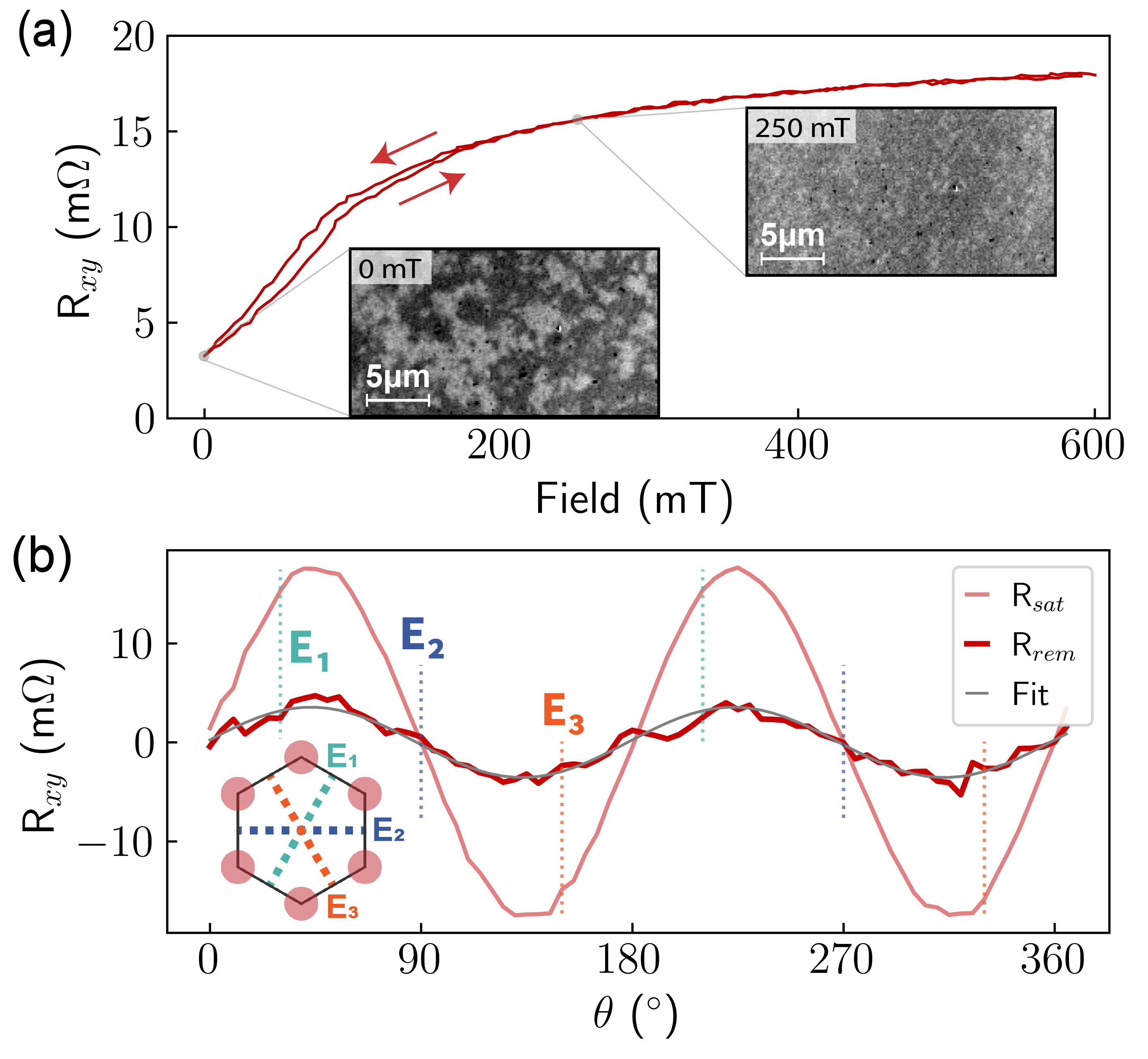}
\caption{(a)~Positive field loop at $45^\circ$. The SMR signal shows small hysteresis below $H_{irr}\approx200$~mT and is fully reversible above $H_{irr}$. Insets show scanned XMLD images of the $\alpha$-Fe$_2$O$_3$ domain structure at 0~mT and 250~mT. Dark/ bright contrast corresponds to horizontal/ vertical orientation of the N\'eel vector. (b)~Transverse remanent resistance (dark red) after $H$ has been relaxed from a saturated state at 0.6~T (light red) to zero field as a function of angle of the applied magnetic field. The inset shows the orientation of the magneto-crystalline easy-axes within the hexagonal unit cell.}
\label{fig:3}
\end{figure}

At room temperature, there is a strong easy-plane anisotropy in $\alpha$-Fe$_2$O$_3$ forcing the magnetic moments to lie in the basal plane.
Additionally, there is a weak six-fold magnetocrystalline anisotropy within this easy-plane with an anisotropy field of $H_{a}\approx1$~$\mu$T~\cite{Tasaki1963,Flanders1965,Besser1967,Chen2012,Cheng2019}. The easy-axes are labeled \textbf{E}$_1$, \textbf{E}$_2$, and \textbf{E}$_3$ in the inset of Fig.~\ref{fig:3}b.

\begin{figure}[ht]
\centering\includegraphics[width=0.85\linewidth]{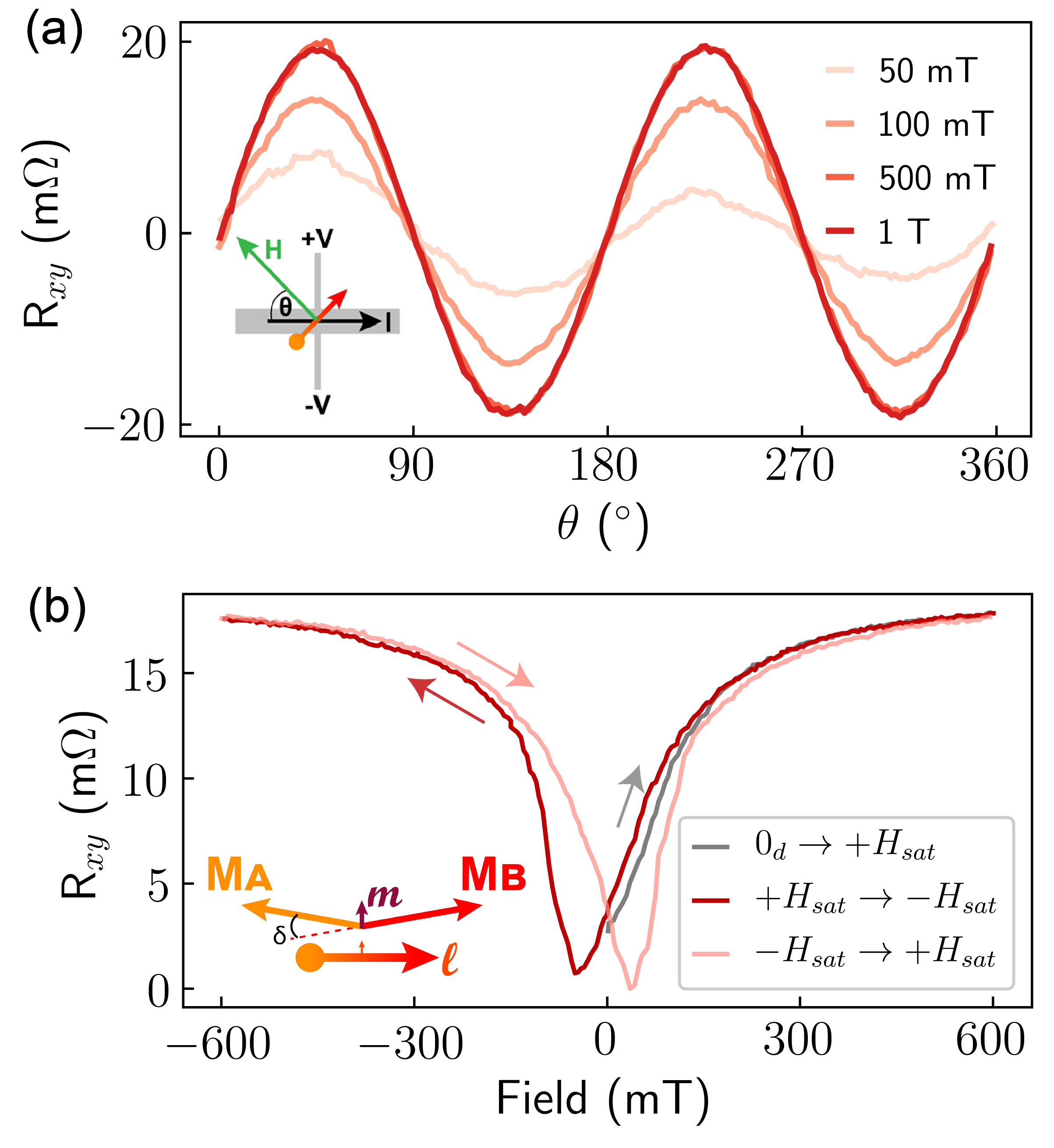}
\caption{(a)~Angle-dependent SMR signal for magnetic field varying from 50 mT (light orange) to 1 T (dark red). Inset shows a schematic of the SMR measurement geometry. (b)~SMR signal as function of field at $\theta=45^\circ$. The gray line shows the initial field sweep from a demagnetized state $0_d$ to $+H_{sat}$, the dark and light red lines show the hysteresis between $+H_{sat}\rightarrow -H_{sat}$ and $-H_{sat}\rightarrow +H_{sat}$ field sweeps, respectively. The inset illustrates the canting of the sublattice magnetizations M$_A$ and M$_B$.}
\label{fig:2}
\end{figure}

The two magnetic sublattices $M_A$ and $M_B$ of $\alpha$-Fe$_2$O$_3$ are slightly canted due to the Dzyaloshinskii-Moriya interaction (DMI)~\cite{Dzyaloshinsky1958, Moriya1960} with a canting angle $\delta=0.13\pm0.01^\circ$\cite{Morrish1995}, giving rise to a small FM moment $m$ in the basal plane of the hexagonal cell (see inset of Fig.~\ref{fig:2}b).
Moreover, $\alpha$-Fe$_2$O$_3$ exhibits pronounced magnetoelastic coupling with a magnetostriction constant $\lambda \approx 10^{-5}$ originating from spontaneous strain~\cite{Urquhart1956,Voskanyan1968b}.\\

The AFM state is commonly read electrically by measuring the angle-dependent spin Hall magnetoresistance (SMR) in $\alpha$-Fe$_2$O$_3$/Pt bilayers where the perpendicular alignment of the N\'eel vector with respect to the external magnetic field results in a negative sign of the SMR signal~\cite{Hoogeboom2017,Fischer2018,Lebrun2019,Geprags2020a}. The electrical measurements presented in this work were performed by applying an AC probing current $I_p = 500 \mu$A and measuring the transverse resistance $R_{xy}$ of the Pt (5~nm thick and 20~$\mu$m wide) crossbar grown on $\alpha$-Fe$_2$O$_3$ by magnetron sputtering.

As shown in Fig.~\ref{fig:2}a, the SMR shows the characteristic $\sin(2\theta)$ dependence, where $\theta$ denotes the angle between the current direction (along easy-axis \textbf{E}$_2$) and the external in-plane magnetic field. A magnetic field smaller than 500~mT is not sufficient to fully reorient the N\'eel order perpendicular to the field, resulting in a smaller amplitude of the angle-dependent SMR.

Fig.~\ref{fig:2}b shows a measurement of the field dependence of the SMR at a fixed angle $\theta=45^\circ$.
Starting from a demagnetized state at zero field $0_d$, $R_{xy}$ initially increases linearly with magnetic field and saturates at $H_{sat} = \pm600$~mT. Above $H_{sat}$, the N\'eel order is oriented fully perpendicular to the magnetic field. When the field is reduced back to zero, the SMR signal decreases to nearly the initial value at $0_d$, signaling that the sample has demagnetized into a multi-domain state with nearly zero net N\'eel vector. This implies the existence of strong internal fields driving the formation of domains.
We have confirmed the strong dependence of the AFM order on the magnetic field by measuring x-ray magnetic linear dichroism (XMLD) contrast images of the domain structure of $\alpha$-Fe$_2$O$_3$ by total electron yield using a scanning x-ray microscope (Maxymus) (see Fig.~\ref{fig:3}a).\\

In contrast to collinear AFMs such as NiO, we observe clear hysteresis in the SMR signal in $\alpha$-Fe$_2$O$_3$/Pt. We define the location of the minima of $R_{xy}$ to be the coercive field $H_c \approx 50$~mT. The hysteresis is a signature of the weak ferromagnetic nature of $\alpha$-Fe$_2$O$_3$. In collinear AFMs with an applied field greater than the spin-flop field, the N\'eel vector can adopt either of the two directions orthogonal to the field. In canted AFMs, on the contrary, the small FM moment prefers to align with the external magnetic field, and hence the $180^\circ$ degeneracy of the orientation of the N\'eel order is lifted due to the coupling of the AFM and FM moments by DMI.
As a consequence, the interpretation of angle-dependent SMR at low fields, where hysteresis plays a significant role, is not straightforward in $\alpha$-Fe$_2$O$_3$.

Therefore, we focus on a field regime with a single polarity, where all N\'eel vectors are oriented within a semicircle defined by the direction of the preceding saturating magnetic field.
Fig.~\ref{fig:3}a shows the SMR signal for an in-plane field sweep from $+H_{sat}$ to zero and back to $+H_{sat}$ oriented at $\theta=45^\circ$.
$R_{xy}$ initially increases linearly with magnetic field. This linear dependence is in stark contrast to the experimentally observed and theoretically modeled quadratic dependence on the magnetic field in the low-field regime previously reported in collinear AFMs such as NiO~\cite{Fischer2018}.
We note that there is a small hysteresis below the irreversibility field $H_{irr}\approx200$~mT. At low fields $0\leq H<H_{irr}$, the reorientation of the N\'eel order is dominated by domain wall movement. Pinning of the AFM domain walls induces irreversibility which manifests in the slight hysteresis of the SMR signal. We can extract an effective pinning field of the AFM domain walls of $H_{p}\approx 20$~mT.
For $H>H_{irr}$, $R_{xy}$ is independent of the preceding magnetic field. In this regime, the reorientation of the N\'eel order is dominated by coherent rotation of the N\'eel vectors within the domains and is, hence, fully reversible.

To elucidate further details of the equilibrium AFM domain structure, we have investigated the angular dependence of the remanent state. We applied an external in-plane magnetic field $H = 1.2$~T~$>H_{sat}$ (light red curve in Fig.~\ref{fig:3}b) and measured the remanent SMR signal $R^{rem}_{xy}$ (dark red curve) as a function of the angle~$\theta$ of the preceding field.
If magneto-crystalline anisotropy were the driving mechanism for relaxation in this system, one would expect the N\'eel vector to relax towards the closest anisotropy axis depending on the direction of the preceding magnetic field. This would result in maxima and minima in R$^{rem}_{xy}$ along \textbf{E}$_1$ and \textbf{E}$_3$ respectively. However, $R^{rem}_{xy}$ does not show any signature of the crystal symmetry.
Fitting $R^{rem}_{xy}$ to $\sin2\theta$ (gray) shows good agreement, implying a simple scaling $R^{rem}_{xy}(\theta)=wR^{sat}_{xy}(\theta)$ with $w\approx0.2$ between the two curves. This suggests that the remanent signal arises from a small fraction of domains that are oriented perpendicular to the preceding saturating magnetic field direction.
\begin{figure}[ht]
\centering\includegraphics[width=0.99\linewidth]{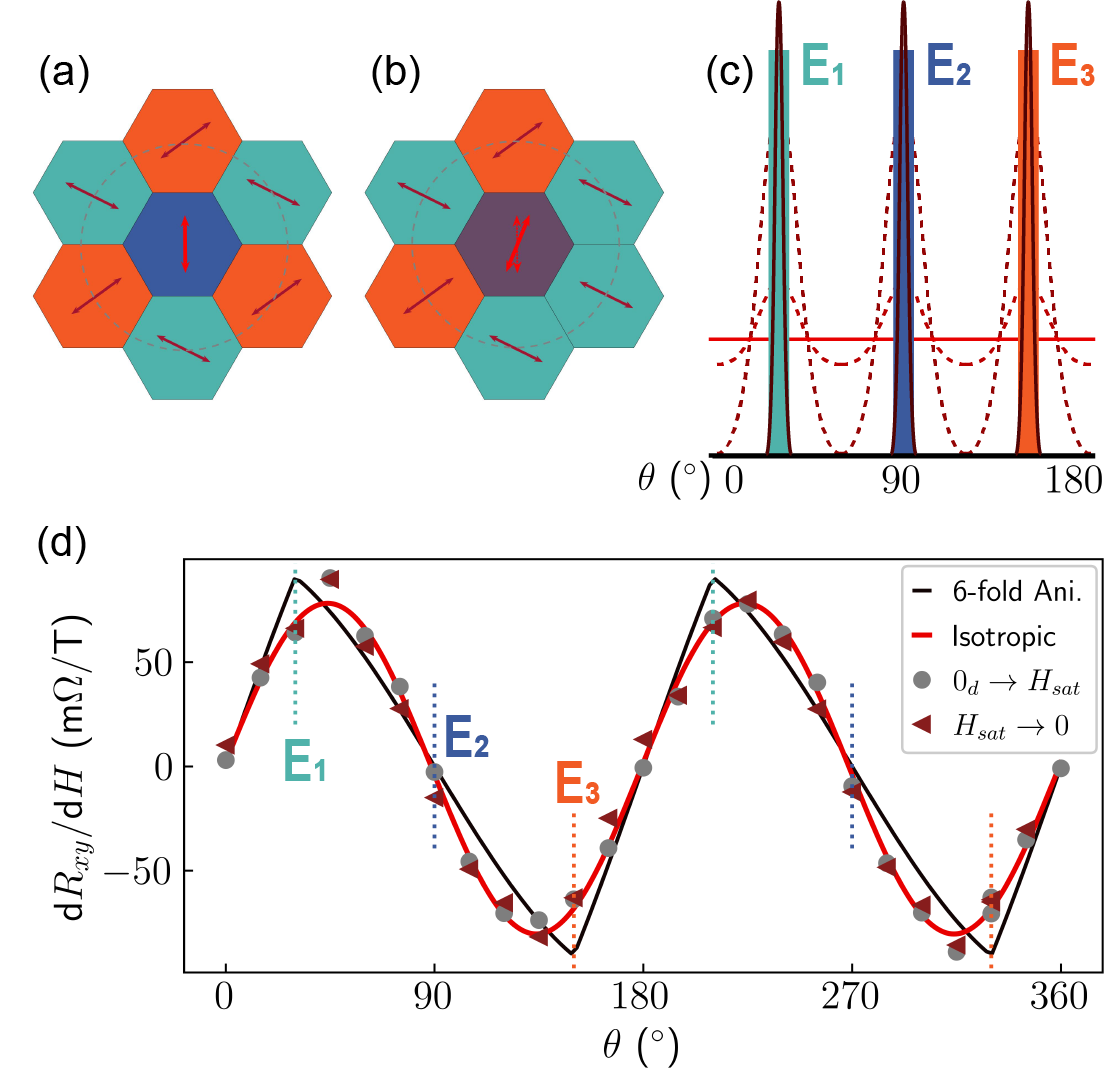}
\caption{(a,b)~Schematic illustrating the deflection of the anisotropy axis away from the magnetocrystalline easy-axis in the central domain depending on the local surroundings. (c)~Schematic of distribution of orientation of N\'eel vectors ranging from strongly aligned with a magnetocrystalline anisotropy (dark gray) to isotropic (red). (d)~Slope of the linear SMR signal at small $H$ as a function of angle of the field calculated using the modified destressing model based on a six-fold anisotropy (black line) and isotropic distribution of N\'eel vector orientations (red line). The isotropic model shows good agreement with the experimental data for a field sweep from a demagnetized state and saturated state (dots and triangles respectively).}
\label{fig:4}
\end{figure}

\section{Theoretical discussion}
In order to identify the different contributions to the formation of the domains, we have investigated the underlying driving mechanisms in more detail.
The formation of a multi-domain state when the magnetic field is reduced has been investigated recently in several AFMs~\cite{Han2014,Hoogeboom2017,Baldrati2018,Fischer2018,Hajiri2019,Zhang2019,Fischer2020}. So far, it has been presumed that the domains are oriented along the magneto-crystalline anisotropy axes. 
The hysteresis and formation of well-defined small domains in zero external magnetic field are evidence for the existence of anisotropy in our system. The lack of signatures of the six-fold symmetry of the magnetocrystalline anisotropy despite the high-quality crystallinity of the epitaxially grown films, however, implies that different contributions are dominant for the effective anisotropy. In the following section, we show that the effective anisotropy is composed of a competition of the short-range magnetocrystalline anisotropy and a long-range contribution of magnetoelastic nature.

First, we note that the spontaneous magnetoelastic strains in the AFM layer are incompatible with the non-strained nonmagnetic substrate and create a long-range so-called destressing field which is similar to  the stray field in ferromagnets~\cite{Gomonay2007,Gomonay2014} (see supplemental material for details ~\cite{SI}). The local orientation of the easy magnetic axes is parameterized by the angle $\varphi(\mathbf{r})$ that is calculated from the easy-axis \textbf{E}$_2$ within the film plane.  By minimizing the energy of the antiferromagnetic layer, we get the following equation for $\varphi(\mathbf{r})$:
\begin{eqnarray}\label{eq_effective equation}
    &&H_\mathrm{an}\sin6\varphi +H_\mathrm{destr}\left[\psi_1\sin 2\varphi-\psi_2\cos 2\varphi\right]=0,\nonumber\\
   &&\psi_1(\mathbf{r})\equiv \frac{1}{4\pi}\int \frac{\cos2\varphi(\mathbf{r }^\prime)}{|\mathbf{r}-\mathbf{r}^\prime|}\delta^\prime(z)d\mathbf{r}^\prime,\\ &&\psi_2(\mathbf{r})\equiv \frac{1}{4\pi}\int \frac{\sin2\varphi(\mathbf{r }^\prime)}{|\mathbf{r}-\mathbf{r}^\prime|}\delta^\prime(z)d\mathbf{r}^\prime, \nonumber
\end{eqnarray}
where $H_\mathrm{an}$ denotes the in-plane magnetic anisotropy and $H_\mathrm{destr}$ is the destressing field which is proportional to the magnetoelastic constant~$\lambda$, and $\delta^\prime(z)$ is the derivative of the Dirac delta function. Here, we neglect the contributions from the AFM domain walls within the plane.

In the second step, we analyse the possible distribution of the effective anisotropy by solving Eq.~(\ref{eq_effective equation}). The first term in Eq.~(\ref{eq_effective equation}) stems from the six-fold crystallographic anisotropy and turns to zero for $\varphi_0=0,\pi$ (direction along \textbf{E}$_2$), $\pi/3,4\pi/3$ (\textbf{E}$_1$), or $2\pi/3,5\pi/3$ (\textbf{E}$_3$). The second term originates from the magnetoelastic coupling and depends on the distribution of the magnetic texture in the whole sample. To estimate this term, we note that the destressing fields $\psi_1(\mathbf{r})$, $\psi_2(\mathbf{r})$ satisfy the Poisson equations
\begin{eqnarray}\label{eq_Poisson}
    -\Delta \psi_1&=& \delta^\prime(z)\cos2\varphi(\mathbf{r}),\\-\Delta \psi_2&=&\delta^\prime(z)\sin2\varphi(\mathbf{r}),\nonumber
\end{eqnarray}
in which the functions $\cos2\varphi(\mathbf{r})$, $\sin2\varphi(\mathbf{r})$ play the role of magnetoelastic charges. 
Based on the analogy with electrostatics, we conclude that the crystalline anisotropy defined by the first term in Eq.~(\ref{eq_effective equation}) can only be restored by either a vanishingly small destressing field $H_\mathrm{destr}$ or an equiprobable distribution of domains in which case the average charges in Eq.~(\ref{eq_Poisson}) vanish. However, in AFM thin-films with finite-sized domains, such averaging can only take place by considering a sufficiently large area including enough different domains. On a smaller scale, the local anisotropy is very sensitive to the direct surroundings. A local imbalance between the domains \textbf{E}$_1$, \textbf{E}$_2$, and \textbf{E}$_3$ creates a local nonzero density of charges leading to a rotation of the local anisotropy axis away from the magnetocrystalline anisotropy axis, as illustrated schematically in Fig.~\ref{fig:4}a,b.
In this example, the N\'eel vector in the central domain in Fig.~\ref{fig:4}a is oriented along \textbf{E}$_2$ as long as the areas $S_1$ and $S_3$ of the surrounding domains along \textbf{E}$_1$ and \textbf{E}$_3$ respectively are equal. In this configuration, the magnetoelastic charge $\sin2\varphi$ is fully compensated.\footnote{~The noncompensated charge $\cos2\varphi$ in this case creates a field $\psi_1$ that affects the value of the anisotropy but not the direction.} However, an imbalance  $S_1\ne S_3$ between the areas along \textbf{E}$_1$ and \textbf{E}$_3$ creates a nonzero charge $\sin2\varphi\propto |S_1-S_3|/(S_1+S_3)$ and a related field $\psi_2$. As a result, the equilibrium orientation of the N\'eel vector in the central domain rotates away from the crystalline axis \textbf{E}$_2$ by $\varphi\propto (H_\mathrm{destr}/ H_\mathrm{an})|S_1-S_3|/(S_1+S_3)$ (Fig.~\ref{fig:4}b). Hence, spatially fluctuating magnetoelastic charges result in local anisotropy axes which do not have to coincide with the crystalline anisotropy axes~\cite{Besser1967}. Depending on the relative strength of the crystalline anisotropy $H_\mathrm{an}$ and the fluctuating destressing fields, the domains in an AFM thin-film can range from distributed sharply around the crystalline anisotropy axes up to an isotropic distribution as illustrated in Fig.~\ref{fig:4}c.

Based on our model, we can introduce a characteristic ''Debye radius'' $D$ at which the effect of fluctuations of the magnetoelastic charges is fully screened. To estimate $D$, we  assume that the orientations of different domains are statistically independent. We can model the probability to find a domain with a given orientation $\varphi$ using the  Boltzmann distribution with effective temperature $T_\mathrm{eff}$. In analogy to the Debye model, we estimate the Debye radius $D=t_\mathrm{AFM}\sqrt{T_\mathrm{eff}/(H_\mathrm{destr}M_s)}$, where $t_\mathrm{AFM}$ is the thickness of the AFM layer, and $M_s/2$ is the sublattice magnetization.

We associate the statistical independence of the domain distribution with the process of the magnetic ordering when cooling through the N\'eel temperature. However, on a short length scale (below the Debye radius $D$) temperature-induced redistribution of the domains can be blocked e.g. by domain wall pinning. In this case correlations between the different domain types result in local fluctuations around the equilibrium distribution and rotation of the N\'eel vector away from the crystallographic directions, as explained above.

In order to verify this model, we investigate the response of the domain structure to an external in-plane magnetic field. For this, we have refined the existing theoretical framework~\cite{Gomonay2002,Fischer2018,Geprags2020b}, which was developed for a compensated three-domain collinear AFM. Here, we consider the competition between the destressing energy and Zeeman energy. Starting from an equiprobable domain distribution at zero field, the existing theory predicts that the dominant domain fraction increases as $H^2/(H_{dest}H_{ex})$ (see Eq.~S27 in SI)~\cite{Lebrun2019}. The DMI in $\alpha$-Fe$_2$O$_3$ and the corresponding nonzero FM moment $m$ give rise to an additional Zeeman term proportional to $ mH/(MH_{dest})$. Our estimations show that the linear Zeeman term dominates over the quadratic term for magnetic fields $H<8$~T. This agrees with our experimental findings that the SMR signal depends linearly on magnetic field in the low field regime (see Figs.~\ref{fig:2}b and \ref{fig:3}a).

To calculate the angular dependence of the SMR signal, we add the Zeeman contribution to Eq.~(\ref{eq_effective equation}) (see SI). We consider two limiting cases of the equilibrium domain distribution schematically shown in Fig.~\ref{fig:4}d: i) isotropic distribution of all $\varphi$ values due to destressing effects (red line); ii) sharp peaks corresponding to six-fold crystalline anisotropy (dark grey line). 
Fig.~\ref{fig:4}d shows the results of the theoretical models and the experimental data of the slope $dR_{xy}/dH$ at low field. The experimentally extracted slope d$R_{xy}/$d$H$ of the field-dependent SMR signal in the linear regime (triangles and dots) shows the qualitative $\sin 2\theta$ angle-dependence irrespective of the preceding magnetic field history. This dependence is in agreement with the theoretical prediction
\begin{equation}
R_{xy}\propto (mH/H_\mathrm{destr})\sin 2\theta
\end{equation}
for an isotropic distribution with the maxima located at $\theta=45^\circ$ and $225^\circ$. In contrast, the six-fold anisotropy leads to peaks at $\theta=30^\circ$ and 150$^\circ$.
This further confirms that the orientation of the domains in AFM thin films does not follow the crystal symmetry but lies along isotropically distributed local anisotropy axes due to substrate clamping.

\begin{figure}[bh]
\centering\includegraphics[width=0.99\linewidth]{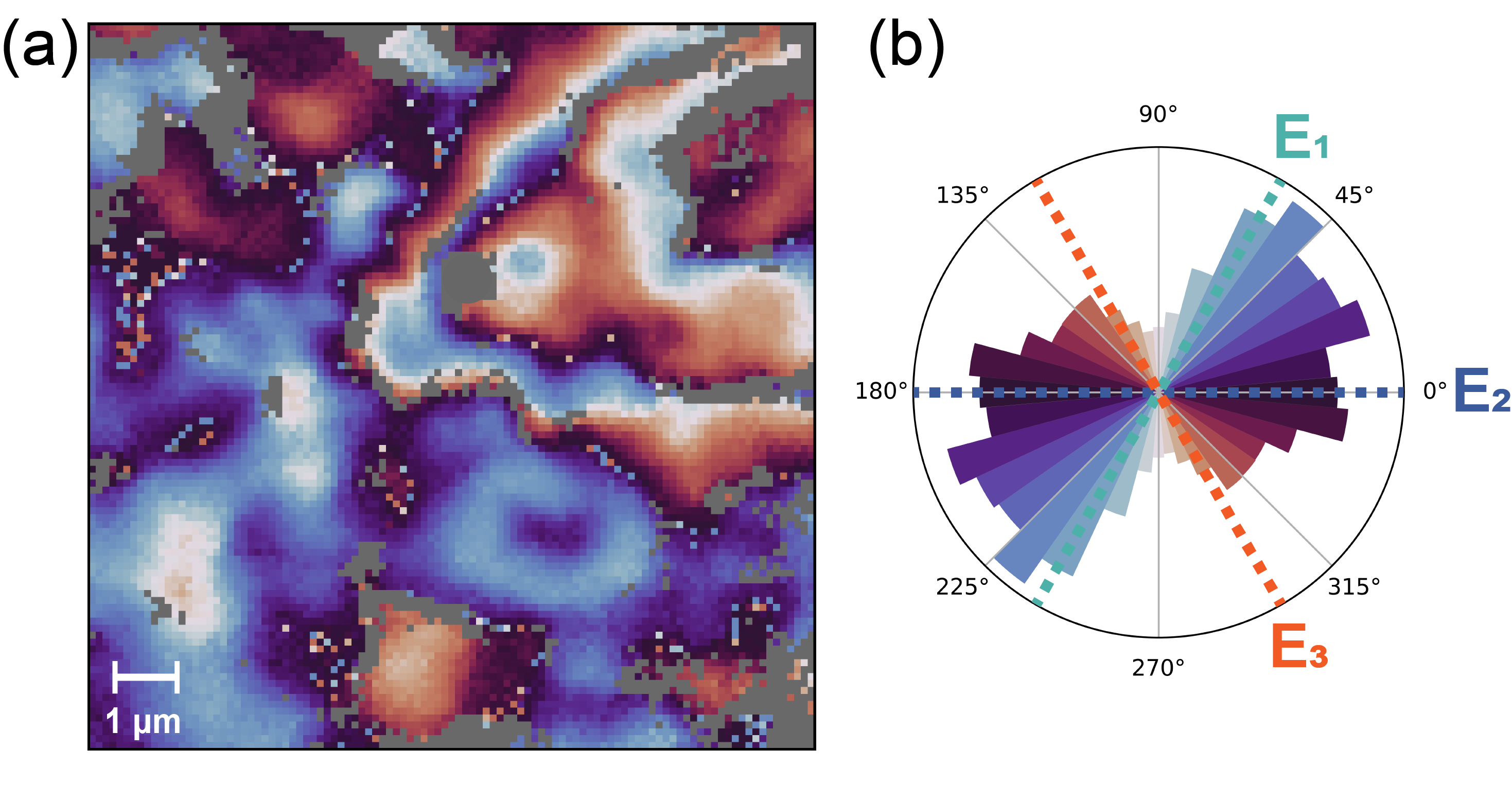}
\caption{(a)~Vector map of the $\alpha$-Fe$_2$O$_3$ domain structure constructed from angle-dependent XMLD XPEEM images. The orientation of the N\'eel vector represented by different colors is shown in the histogram in (b). The height of the bars in the circular histogram mark the frequency of occurrence of the given orientation of the N\'eel vectors on a linear scale. Pixels for which the uncertainty is larger than 15$^\circ$ are marked in gray.}
\label{fig:5}
\end{figure}

\section{Experimental verification}
In order to investigate the orientation of the N\'eel vectors in the multi-domain state, we have performed angle-dependent XMLD x-ray photoemission electron microscopy (XPEEM) on $\alpha$-Fe$_2$O$_3$ (20~nm with 2~nm Carbon capping). Varying the angle between the linear polarization of the incident x-rays and the sample allows reconstruction of the orientation of the N\'eel orientation (see supplemental material for details of analysis ~\cite{SI}).

Fig.~\ref{fig:5}a shows a vector map of the antiferromagnetic domain structure. The orientation of the N\'eel vectors encoded by different colors is shown in the histogram in Fig.~\ref{fig:5}b. While the histogram does not show any signatures of the hexagonal crystal structure of $\alpha$-Fe$_2$O$_3$, the distribution of domain orientation is not fully isotropic either in the limited field of view of the figure. This agrees well with the model of locally varying effective anisotropies we introduce above.

\section{Conclusions}
This work highlights the impact of strain effects on the equilibrium domain structure in thin-film AFMs. Combining imaging of the AFM domain structure with electrical measurements, we find that the formation of domains is predominantly driven by destressing fields which do not follow the hexagonal crystal symmetry of $\alpha$-Fe$_2$O$_3$ but rather result in an overall isotropic distribution of the N\'eel vector orientations with fluctuating easy-axes depending on the local surrounding domains.
Our work demonstrates that understanding the formation of the domain structure and effective anisotropy in thin-films is paramount for realizing AFM-based memory technologies.

\begin{acknowledgements}
The authors thank M. Birch, S. Meyer, D. Bono, and B. Neltner for their valuable contributions to this work. We thank A. Lee for the use of his Raman spectrometer and P. Askeland for technical assistance.
The x-ray imaging measurements were carried out at the UE-49-PGM-SPEEM (XPEEM) and UE46 MAXYMUS (STXM) beamlines at Helmholtz-Zentrum Berlin. We thank HZB for the allocation of synchrotron radiation beamtime.
This work is supported in part by SMART, one of seven centers of nCORE, a Semiconductor Research Corporation program, sponsored by the National Institute of Standards and Technology (NIST), and by the DARPA TEE program. A. W. acknowledges financial support from the Swiss National Science Foundation. O.G. acknowledges support of the ERC Synergy Grant SC2 (No. 610115), the Deutsche Forschungsgemeinschaft (DFG, German Research Foundation) - TRR 173 – 268565370 (project B12) and TRR 288 – 422213477 (project A09). A.L. and D.W. acknowledge funding from the Würzburg-Dresden Cluster of Excellence on Complexity and Topology in Quantum Matter—ct.qmat (EXC 2147, project no. 390858490). F.B. acknowledges financial support from the Helmholtz Young Investigator Group Program.
N. B. thanks G.S.D. Beach and his group for their hospitality during his sabbatical visit to MIT.
\end{acknowledgements}

\bibliographystyle{apsrev4-2}
\bibliography{references} 
\end{document}